\documentclass[prd,twocolumn,eqsecnum,preprintnumbers,showpacs]{revtex4}
\usepackage{amsfonts}
\usepackage{amsmath}
\usepackage{amssymb}
\usepackage{bm}
\usepackage{graphicx}
\usepackage{dcolumn}
\usepackage[latin1]{inputenc}
\usepackage{latexsym}
\usepackage{rotating}
\usepackage{hyperref}

\newcommand{\be}{\begin{equation}}
\newcommand{\ee}{\end{equation}}

\newcommand{\ba}{\begin{eqnarray}}
\newcommand{\ea}{\end{eqnarray}}

\begin{document}

\title[Chern-Simons Modified Gravity as a Torsion Theory  and its
Interaction with Fermions]{Chern-Simons Modified Gravity as a Torsion
Theory and its Interaction with Fermions} 

\author{Stephon Alexander}
\author{Nicol\'as Yunes} 

\affiliation{Institute for Gravity and the Cosmos,
  Department of Physics, The Pennsylvania State University, University
  Park, PA 16802, USA}

\date{\today}

\preprint{IGC-08/4-1}

\begin{abstract}
  
  We study the tetrad formulation of Chern-Simons (CS) modified
  gravity, which adds a Pontryagin term to the Einstein-Hilbert action
  with a spacetime-dependent coupling field. We first verify
  that CS modified gravity leads to a theory with torsion, where this
  tensor is given by an antisymmetric product of the Riemann tensor and
  derivatives of the CS coupling. We then calculate the torsion
  in the far field of a weakly gravitating source within the
  parameterized post-Newtonian formalism, and specialize the result to
  Earth. We find that CS torsion vanishes only if the
  coupling vanishes, thus generically leading to
  a modification of gyroscopic precession, irrespective of
  the coupling choice. Perhaps most interestingly, we couple fermions
  to CS modified gravity via the standard Dirac action and find that
  these further correct the torsion tensor. Such a correction leads to two new results: 
  (i) a generic enhancement of CS modified gravity by the Dirac equation and axial fermion 
  currents; (ii) a new two-fermion interactions, mediated by an axial current and the CS
  correction. We conclude with a discussion of the consequences of these results in particle 
  detectors and realistic astrophysical systems.

\end{abstract}

\pacs{04.50.Kd,04.20.Fy,04.40.Nr,04.60.Cf,04.60.Pp}




\maketitle

\section{Introduction}\label{intro}

A quantum gravitational theory that is mathematically consistent,
predictive and in agreement with all experimental data is one of the
holy grails of physics. Many extensions of General Relativity (GR)
have been proposed since its inception, most of which have not passed
the test of time and increasingly more accurate experiments (see
e.~.g.~\cite{Will:2005va} for a current review). Recently, however,
two competing paradigms have arisen that hold the promise to unify GR
with quantum theory: String
Theory~\cite{Polchinski:1998rq,Polchinski:1998rr,Polchinski:1994mb}
and Loop Quantum
Gravity~\cite{Ashtekar:2004eh,lrr-1998-1,Smolin:2004sx}.

Although both these extensions are technically theoretically 
incomplete, there has been a recent effort to study its
predictability~\cite{Damour:2008ji,Kallosh:2007wm}. Due to the
intrinsic complexity of these theories, such efforts have been
traditionally limited or model dependent~\cite{Mavromatos:2000ym}.
Recently, however, these theories have advanced enough that
predictions can be made and one generic and unavoidable low-energy
limit of both theories has been discovered: Chern-Simons (CS) modified
gravity.

In String Theory, the absence of a CS term in the action leads to the
Green-Schwarz anomaly, which {\emph{requires}} cancellation to
preserve unitarity and quantum consistency. In most perturbative
string theories (e.~g.~Type IIB, I, Heterotic) with four-dimensional
compactifications, the Green-Schwarz mechanism requires the inclusion
of a CS term~\cite{Alexander:2004xd}. In fact, this term is induced in
{\emph{all}} string theories due to duality symmetries in the presence
of Ramond-Ramond scalars or D-instanton
charges~\cite{Polchinski:1998rr,Alexander:2004xd}. Even in heterotic
M-theory the CS term is required through the use of an anomaly inflow.

In Loop Quantum Gravity, the CS term arises as a natural extension to the Hamiltonian constraint. In particular, the CS term renders a candidate Holomorphic ground state  wavefunction invariant under large gauge
transformations of the Ashtekar connection
variables~\cite{Ashtekar:1988sw}. The CS correction, is also related
to the Immirzi parameter of Loop Quantum Gravity, which determines the
spectrum of quantum geometrical operators~\cite{Perez:2005pm,Randono:2005up}. 

CS modified gravity proposes an extension to GR by adding a parity-violating,
Chern-Pontryagin term to the Einstein-Hilbert action, multiplied by a
spacetime-dependent coupling scalar~\cite{Jackiw:2003pm}. This theory
modifies the GR field equations by adding a new Cotton-like C-tensor,
which is composed of derivatives of the Ricci tensor and the dual to the
Riemann. Additionally, the equations of motion for the scalar field
provide a new Pontryagin constraint that preserves diffeomorphism
invariance. The structure of the C-tensor allows the modified theory
to preserve some of the classical solutions of GR, such as the
Schwarzschild, the Friedmann-Robertson-Walker and the gravitational
wave line elements~\cite{Jackiw:2003pm,Guarrera:2007tu}.

Although some classic GR solutions are preserved in CS modified
gravity, parity violation is inherent in the modified theory, leading
to possibly observable effects. One such effect is amplitude
birefringence~\cite{Jackiw:2003pm,Alexander:2004wk}, which leads to a
distinct imprint that could be detectable through gravitational wave
observations~\cite{Alexander:2007kv,LIGO}. Birefringent
gravitational waves have actually been successfully employed to propose
an explanation to the leptogenesis
problem~\cite{Alexander:2004us,Alexander:2007qe} and could also leave
an imprint in the cosmic-microwave
background~\cite{Lue:1998mq,Li:2006ss,Alexander:2006mt}. Another consequence 
of CS modified gravity is modified precession, which has
been studied in the far field
limit~\cite{Alexander:2007zg,Alexander:2007vt}, leading to a weak
bound on the CS scalar with LAGEOS~\cite{Smith:2007jm}. Recent
investigations have also concentrated on spinning black hole
solutions~\cite{Konno:2007ze,Grumiller:2007rv}, as well as black hole
perturbations~\cite{Yunes:2007ss}, both of which have been seen to be
corrected in CS modified gravity. For further studies of these and related issues
see e.g.~\cite{Kostelecky:2003fs,Mariz:2004cv,Bluhm:2004ep,Eling:2004dk,Lyth:2005jf,Mattingly:2005re,Lehnert:2006rp,Hariton:2006zj,Fischler:2007tj,Tekin:2007rn}.
and references therein.

In this paper, we study CS modified gravity within the first-order or
tetrad formalism (see e.~g.~\cite{Romano:1991up} for a review). In
this formalism, one rewrites the action in terms of a tetrad and a
generalized connection that need not be torsion-free. One then varies
the action with respect to these fields to obtain the equations of motion and the so-called
second Cartan structure equation, which in GR reduces to the torsion-free
condition. CS modified gravity, however, leads to a
torsion-full condition, where the torsion tensor is proportional to an antisymmetric 
product of the Riemann tensor and partial derivatives of the CS scalar.

We first compute the torsion tensor in the far field of a weakly
gravitating body within the parameterized post-Newtonian (PPN)
formalism for a generic CS scalar~\cite{Schiff:1960gi,Nordtvedt:1968qs,1972ApJ...177..775N,1971ApJ...163..611W,1973ApJ...185...31W,Will:1993ns}. We find that the torsion tensor is
proportional to contractions of the Levi-Civita symbol, derivatives of
the CS scalar and derivatives of the Newtonian and PPN vector
potentials. This tensor is evaluated around Earth and found to
generically persist, unless the CS scalar field vanishes identically, thus reducing CS
modified gravity to GR. The non-vanishing of the CS torsion
tensor generically leads to a modified frame-dragging effect and
gyroscopic precession. The results found here thus 
provide great theoretical motivation for studies of generic
torsion theories and their effect in Solar System experiments similar to~\cite{Mao:2006bb}.

After investigating the torsion tensor, we concentrate on the
inclusion of fermions in CS modified gravity, since these are known to
also lead to torsion (see e.~g.~\cite{Mercuri:2007ki,Perez:2005pm}). We find that
indeed the torsion tensor is now given by the sum of the CS torsion and
a new fermion-induced term, which depends on the axial fermion 
current. The fermion-extended torsion tensor can then be used to obtain two new results: 
a new two-fermion interaction and a fermionic enhancement of CS modified gravity. 

Interaction terms are common in torsion-full theories. For example, Riemann-Cartan theory
leads to a four-fermion interaction term, mediated by the axial current. These interactions are computed by inserting into the action the full connection: a torsion-free, symmetric part (the Christoffel connection) plus a certain linear combination of components of the torsion tensor (the contorsion tensor). In the fermion-extended version of CS modified gravity, we find that the interaction term consists of three contributions: a new two-fermion term, a modified four-fermion term, and a new six-fermion term. The four-fermion interaction is in fact similar to that found in Riemann-Cartan theory, also suppressed by a factor of $G$, the gravitational constant. The six-fermion interaction is further suppressed by a factor of $G^{2}$. The two-fermion process, however, is $G$-independent and mediated by derivatives of the CS scalar, the axial fermion current, the Ricci scalar and the Ricci tensor. 

The fermion enhancement effect arises as a consequence of the Dirac equation in fermion-extended CS modified gravity. Due to the inclusion of fermions, a new field equation
arises (the Dirac equation), which couples derivatives of the Dirac spinor to the connection, which now contains both a symmetric, torsion-free part and a torsion-full piece. In this way, the torsion tensor, and thus, the CS correction,  are {\emph{sourced}} by derivatives of the Dirac spinor through the Dirac equation. Such a result implies that all CS corrections are magnified in physical scenarios where fermionic currents are large. 

We conclude with a discussion of the consequences of these two new results. On the one hand, the new two-fermion interaction could potentially lead to observables, related to fermion processes. Particle accelerators, however, are unlikely to see this correction, since the Ricci scalar vanishes in the neighborhood of the Solar System, thus annihilating the modification.  On the other hand, the fermionic enhancement effect renders the modified theory even more appealing, since CS corrections would then be naturally enhanced in several realistic astrophysical scenarios, such as pulsars, merging neutron stars, and supernovae, perhaps even leading to stronger bounds of CS modified gravity. 

The remaining of this paper presents further details and calculations
of the results mentioned above and it is divided as follows:
Sec.~\ref{1st-GR} reviews the tetrad formalism in GR and establishes
notation; Sec.~\ref{1st-CS} reformulates CS modified gravity in the
tetrad formalism and finds the torsion tensor of the modified theory;
Sec.~\ref{Torsion-far-field} computes the torsion tensor in the far
field of a weakly gravitating body, later specializing the result to
fields around Earth; Sec.~\ref{Adding-Fermions} adds fermions to the
modified theory, derives the fermionic enhancement effect and calculates the new fermion interactions; Sec.~\ref{Discussion} discusses the implications of these results in astrophysical scenarios and particle detectors; Sec.~\ref{Conclusions}
concludes and points to future research.

We use the following conventions: commas stand for partial derivatives
$\partial_a \psi = \psi_{,a}$; parenthesis and square-brackets in
index lists stand for symmetrization $A_{(ab)} = 1/2 (A_{ab} +
A_{ba})$ and antisymmetrization $A_{[ab]} = 1/2 (A_{ab} - A_{ba})$
respectively; upper-case Latin letters $\{A,B,\ldots\}$ stand for
internal indices, lower-case Latin letters at the beginning of the
alphabet $\{a,b,\ldots, h\}$ stand for spacetime indices, while those
in the middle of the alphabet $\{i,j,\ldots\}$ stand for spatial
indices only. The order symbol ${\cal{O}}(A)$ stands for terms of
order $A$ and we use geometric units, such that $G = 1 = c$.

\section{First Order Formalism in GR}
\label{1st-GR}

In this section, we review the first order formalism of GR and establish notation, following
mainly~\cite{Romano:1991up}. Let us then consider a $4$-dimensional
manifold ${\cal{M}}$ with an associated $4$-dimensional metric
$g_{ab}$. Let us further introduce at each point on the manifold a tetrad $e_a^I$, such that the metric can be written as $g_{ab} = e_a^I e_b^J \eta_{IJ}$, with $\eta_{IJ}$ the Minkowski metric.
Internal and spacetime indices are raised and lowered with $\eta_{IJ}$ and $g_{ab}$ respectively. 



Let us now introduce the spacetime and spin connection, $A_{ab}{}^c$ and
$A_{aI}{}^J$, which given a mix tensor $g_{bI}$ satisfy
\be
D_{a} k_{bI} = \partial_a k_{b I} + A_{ab}{}^{c} k_{c I} + A_{a I}{}^J
k_{b J},
\ee
where $D_a$ is a generalized covariant derivative. 
The torsion tensor is defined via  $2 D_{[a} D_{b]} f = T_{ab}{}^{e} D_{e} f$, 
for some scalar function $f$, thus satisfying 
\be
\label{def-torsion}
T_{ab}{}^e := 2 A_{[ab]}{}^e.
\ee
The requirement that the spin connection be torsion-free is simply $A_{[ab]}{}^c =
0$ and that it  be compatible with the internal metric $\eta_{IJ}$ is equivalent to
$A_{a(IJ)} = 0$. 

The generalized covariant derivative can be shown to be compatible with the tetrad, thus satisfying
\be
D_{a} e_{b}^{I} = 0.
\ee
This relation then implies that the spacetime and spin connections are related via
\be
\label{chg-basis}
A_{ab}{}^{e} = \left(e^{I}_{e}\right)^{-1} A_{aK}{}^{I} e_{b}^{K} - \left(e_{e}^{I}\right)^{-1} \partial_{a} e_{b}^{I},
\ee
which is simply a change of basis. Sometimes these relations are referred to as the ``tetrad postulate,'' which we discuss further in the Appendix. When the spacetime and spin connections satisfy Eq.~\eqref{chg-basis}, then the spacetime connection is given by the sum of the Christoffel symbols and the contorsion tensor (provided the spin connection is torsion-full). The contorsion tensor shall be discussed later, but it is essentially constructed from the torsion tensor. 

With this generalized covariant derivative and connections we can
now define the generalized curvature tensors through the
failure of commutativity of the generalized covariant derivatives. One
can show that
\begin{subequations}
\ba
F_{abI}{}^J &=& 2 \partial_{[a} A_{b] I}{}^J +
\left[A_a,A_b\right]_I{}^J,
\\
F_{abc}{}^d &=& 2 \partial_{[a} A_{b] c}{}^d +
\left[A_a,A_b\right]_c{}^d,
\ea
\end{subequations}
where the anticommutator is short-hand for 
\ba
\left[A_a,A_b\right]_I{}^J &:=& A_{aI}{}^K A_{bK}{}^{J} - A_{bI}{}^K
A_{aK}{}^{J},
\nonumber \\ 
\left[A_a,A_b\right]_c{}^d &:=& A_{ac}{}^e A_{be}{}^{d} - A_{bc}{}^e
A_{ae}{}^{d}. 
\ea
Note that if the connection is metric compatible and torsion-free (i.~e.~if it is the Christoffel connection), then the curvature tensor is simply the Riemann tensor.

Let us now rewrite the Einstein-Hilbert action in terms of these new
variables. Note, however, that we wish to work with the trace of the
generalized curvature tensor, and not the Ricci scalar, since these
two quantities are not necessarily equivalent in the presence of torsion. 
The Einstein-Hilbert action is given by the well-known expression 
\be
\label{S_EH}
S_{EH} = \frac{\kappa}{4} \int d^4x \;\; \tilde{\eta}^{abcd}
\epsilon_{IJKL} e_a^I e_b^J F_{cd}{}^{KL},
\ee
where $\tilde{\eta}^{abcd}$ is the Levi-Civita symbol and
$\epsilon_{abcd} = e^I_a e^J_b e^K_c e^L_d \epsilon_{IJKL}$ is the
Levi-Civita volume form.  We here depart slightly from the conventions of~\cite{Romano:1991up} by not adding an extra factor of two in the action, which is a matter of convention. Equation~\ref{S_EH} can be derived by using the identity
\be
F = \delta^{b}_{[d} \delta^c_{e]} F_{bc}{}^{de},
\ee
and~\cite{Carroll:2004st}
\begin{subequations}
\ba
\epsilon^{a b c d} \epsilon_{a b e f} &=& - 4 \; \delta^{[c}_e
\delta^{d]}_f, 
\\
\tilde{\eta}^{a b c d} \epsilon_{a b e f} &=& + 4 \; \sqrt{-g} \; 
\delta^{[c}_e \delta^{d]}_f,  
\\
\tilde{\eta}^{a b c d} \tilde{\eta}_{a b e f} &=& + 4 \; \delta^{[c}_e
\delta^{d]}_f,   
\ea
\end{subequations}
which will be extremely useful in later section.

Let us now obtain the field equations of the theory by varying the
Lagrangian density with respect to the tetrad and the connection: 
the field equations and the second Cartan
structure equation. Variation with respect to the tetrad
yields
\be
\label{1st-Cartan}
\tilde{\eta}^{abcd} \epsilon_{IJKL} e^J_b F_{cd}{}^{KL} = 0,
\ee
since the curvature tensor depends only on the connection.
Equation~\eqref{1st-Cartan} constitutes the field equations,
which is a generalization of the Einstein field equations for a generic, not necessarily torsion-free 
connection.

Variation with respect to the connection is a bit more complicated.
Let us begin by rewriting the variation of the curvature tensor as
\be
\delta F_{cd}{}^{KL} = 2 D_{[c} \delta A_{d]}{}^{KL} - T_{cd}{}^{e} \delta A_{e}{}^{KL}. 
\ee
Before we vary this Lagrangian density with respect to the connection, it is convenient to
integrate by parts the first term to find
\ba
 \delta S_{EH} &=& - \frac{\kappa}{4} \int d^4x  \; \left[ 2 D_{[c}
  \left(\tilde{\eta}^{abcd} \epsilon_{IJKL} e_a^I e_b^J \right) \delta
  A_{d]}{}^{KL}  
\right.
\nonumber \\
&+& \left. 
T_{cd}{}^e \tilde{\eta}^{abcd} \epsilon_{IJKL} e_a^I e_b^J \delta
A_e{}^{KL} \right].
\ea
We can now vary this action with respect to $A_{e}{}^{KL}$ and demand that the variation vanishes to find
\be
\label{2nd-structure}
- 2 D_{c} \left(\tilde\eta^{abce} \epsilon_{IJKL }e_{a}^I e_b^J \right) =  T_{cd}{}^e \tilde\eta^{abcd} \epsilon_{IJKL} e_a^I e_b^J.
\ee
The left-hand side of this equation vanishes because the generalized covariant
derivative is tetrad compatible and thus Eq.~\eqref{2nd-structure} is simply the torsion-free condition of
GR, 
\be
T_{cd}{}^{e} = 0.
\ee
\%
In this case, then, the
generalized connection reduces to the Christoffel one and the field
equations to the Einstein equations.

\section{First Order Formalism Of CS Modified Gravity}
\label{1st-CS} 

In this section we shall present a pedagogical introduction to CS modified 
gravity in the second-order formalism and derive its first order former. This section
will thus both serve as an introduction to the modified theory, which was originally proposed
in second-order form, and as a basis to establish the CS notation of this paper. 

CS modified gravity~\cite{Jackiw:2003pm} postulates the
following action~\footnote{We here define the CS
  correction with a minus sign, correcting a typo
  in~\cite{Grumiller:2007rv}} 
\begin{subequations}
\ba
S &=& S_{EH} + S_{CS}
\\
\label{CS-S}
S_{CS} &=& \kappa \int d^4x \sqrt{-g} \left(+ \frac{1}{4} \theta {}^{\star}RR
\right),
\ea
\end{subequations}
where $S_{EH}$ is given in Eq.~\eqref{S_EH} and we follow the
conventions of~\cite{Grumiller:2007rv}.  The quantity $\theta$ is here the
so-called {\emph{CS scalar}}, which serves as a spacetime
coordinate-dependent coupling function. In principle, one should include a kinetic and a potential term for the scalar field in the CS action, but we shall ignore these here since they do not contribute to torsion. The Chern-Pontryagin term is
defined via
\be
{}^{\star}RR := \frac{1}{2} \epsilon^{cdef} R^a{}_{bef} R^b{}_{acd}, 
\ee
with $R_{abcd}$ the Riemann tensor. The parity-violating nature of CS
modified gravity is encoded in the Levi-Civita tensor. Note here that CS modified gravity is intrinsically
$4$-dimensional, which is different from the $2+1$-dimensional theory
that goes by a similar name. 

Before decomposing CS modified gravity in first order form, it is convenient to slightly rewrite the action. Let us then integrate by parts to obtain
\be
S_{CS} = - \frac{\kappa}{2} \int d^4 x \sqrt{-g} v_{a} K^a.
\ee
We here neglect any boundary terms since~\cite{Grumiller:2008ie} has shown, 
within the second order formulation, that CS modified gravity indeed leads to a well-posed boundary
value problem, through the addition of boundary counter-terms.
The {\emph{CS velocity}} and {\emph{CS acceleration}} are defined via
\begin{subequations}
\ba
v_a &:=& \nabla_a \theta = \partial_a \theta,
 \\
v_{ab} &:=& \nabla_a v_b = \nabla_a \nabla_b \theta,
\ea
\end{subequations}
where $\nabla_a$ is the covariant derivative operator associated with
the Christoffel connection $\Gamma_{ab}^c$. The quantity $K^a$ is
the so-called Pontryagin current, which in four-dimensions is given by
\be
K^a := \epsilon^{abcd} \Gamma_{bf}{}^e \left( \partial_c \Gamma_{de}{}^f +
  \frac{2}{3} \Gamma_{ce}{}^l  \Gamma_{dl}{}^f\right),
\ee
and satisfies $\nabla_a K^a = {}^{\star}RR/2$~\footnote{This
  corrects a typo in Eq.~(7) of~\cite{Grumiller:2007rv}, which does not
  affect the results of that paper}.

We can now write the CS action in first-order form. The Pontryagin current
can be written in terms of the Riemann tensor as 
\be
\label{R-current}
K^a =  \epsilon^{abcd} \Gamma_{bf}{}^e \left( \frac{1}{2} R_{cde}{}^{f} -
  \frac{1}{3} \Gamma_{ce}{}^l \Gamma_{dl}{}^f \right),
\ee
where we again follow the convetions of~\cite{Romano:1991up} and define the Riemann tensor via $R_{cde}{}^{f}= 2 \partial_{[c} \Gamma_{d]e}{}^{f} + 2 \Gamma_{[c|e}{}^{l} \Gamma_{d]l}{}^{f}$. 
We then find that the CS action in first order form is simply
\ba
S_{CS} &=& \frac{\kappa}{2} \int d^4x\;\;  \tilde{\eta}^{abcd} v_a
A_{bI}{}^{J} \left[ \frac{1}{2} F_{cdJ}{}^I - \frac{1}{3} A_{cJ}{}^K
  A_{dK}{}^I \right],
  \nonumber \\
\label{1stCS}
\ea
where we used that $\epsilon^{abcd} = (-1/\sqrt{-g}) \;
\tilde{\eta}^{abcd}$ and we have replaced spacetime by internal indices, 
since the these are fully contracted. 



Let us now vary the first-order CS action with respect to the tetrad 
and the connection. The field equations remain
the same as in GR, namely Eq.~\eqref{1st-Cartan}, because the CS
action does not depend on the tetrad. We then find that
the field equations of CS modified gravity are similar to those of GR,
provided the connection and the curvature tensor are the generalized ones.

The second structure equation is a bit more difficult to
derive. Let us then first perform a general variation of  
the CS modified action in first-order form to
find
\be
\delta S = \frac{\kappa}{4} \int d^4x  v_a \tilde{\eta}^{abcd}
\left( \delta A_{bI}{}^{J} F_{cdJ}{}^I - \delta E \right),
\ee
where 
\ba
\delta E &=&  A_{bI}{}^{J} \delta F_{cdJ}{}^I - \frac{2}{3}
  \delta A_{cJ}{}^K A_{dK}{}^I A_{bI}{}^{J} 
\\ \nonumber 
&-& 
\frac{2}{3} A_{cJ}{}^K
  \delta A_{dK}{}^I  A_{bI}{}^{J}
- \frac{2}{3} \delta A_{bI}{}^{J} 
  \delta A_{cJ}{}^K A_{dK}{}^I.
\ea
Upon variation with respect to $A_{a}{}^{KL}$ and contraction with the
Levi-Civita symbol, the above term identically vanishes and we are left with 
\be
\label{var-S}
\frac{\delta S_{CS}}{\delta A_{a}{}^{KL} } = 
\frac{\kappa}{4} \int d^4x \;
\tilde{\eta}^{abcd}  \;  v_b F_{cdKL}.
\ee
Combining Eq.~\eqref{var-S} with the variation of the Einstein-Hilbert action with respect to the spin connection we find the second structure equation, namely 
\be
\label{2nd-cartan}
\epsilon_{IJKL} \tilde\eta^{abcd} T_{cd}{}^e e_a^I e_b^J =  \tilde \eta^{ebcd} v_b F_{cdKL}, 
\ee
which agrees with~\cite{Cantcheff:2008qn} up to conventional prefactors. 

Let us now attempt to isolate the torsion tensor in CS modified
gravity. Equation~\eqref{2nd-cartan} is in principle a differential equation for the torsion 
tensor, since the generalized curvature tensor contains
derivatives of the contorsion. 
Following~\cite{Perez:2005pm}, we can parameterize the full connection via 
\be
\label{full-connection}
A_{a}{}^{IJ} = \omega_{a}{}^{IJ} + C_{a}{}^{IJ},
\ee
where $\omega_{a}{}^{IJ}$ is a torsion-free, symmetric connection that depends only on the tetrad and $C_{a}{}^{IJ}$ is the contorsion tensor. The contorsion is related to the torsion via
\be
\label{conteq1}
T_{ab}{}^{c} e_{cI} = 2 C_{[a}{}^{IK} e_{b] K},
\ee
where the factor of two comes from our definition of the torsion tensor 
(see Appendix~\ref{app-cont}). We can thus schematically rewrite Eq.~\eqref{torsion-eq1} as
\ba
\label{torsion-eq2}
\epsilon_{IJKL} \tilde{\eta}^{abcd}T_{cd}{}^e e_a^I e_b^J  &=& \tilde{\eta}^{ebcd} v_b R_{cdKL}[\omega] 
\\ \nonumber 
&+& \tilde{\eta}^{ebcd} v_b H_{cdKL}[\partial T, \omega T, T^{2}].
\ea
where $R_{abIJ}[\omega]$ is the standard Riemann curvature tensor that depends on $\omega_{a}{}^{IJ}$ only, while $H_{abIJ}[\partial T,\omega T, T^{2}]$ represents all other terms in the generalized curvature tensor that are at least linear in the torsion tensor. The solution to this equation to linear order in the CS velocity is simply 
\ba
\label{torsion}
T_{cd}{}^n &=& - \frac{1}{4}  \epsilon^{nbef} v_b R_{cdef} + {\cal{O}}(v)^{2},
\nonumber \\
T_{cd}{}^{n} &=& - \frac{1}{4} \; {}^{\star}R_{cd}{}^{nb} \; v_{b}  + {\cal{O}}(v)^{2}.
\ea
One can easily check that inserting $T_{cd}{}^{n} = {}^{(2)}T_{cd}{}^{n}$ into Eq.~\eqref{torsion-eq2}, where $^{(2)}T_{cd}{}^{n} = {}^{(1)}T_{cd}{}^{n} + \zeta$ ,  $^{(1)}T_{cd}{}^{n}$ is the first order solution given in Eq.~\eqref{torsion} and $\zeta$ is undetermined, forces $\zeta$ to be at least quadratic in $v_{a}$.  
\section{CS Torsion in the Far Field}
\label{Torsion-far-field} 

The torsion tensor found in the previous section has an intriguing
form, resembling the wedge-product of the Riemann tensor and the CS
velocity. In this section we study the structure of the torsion tensor
in the far field. We begin by considering its functional form in the
PPN formalism and finish with a discussion of this tensor around Earth.

\subsection{CS Torsion in the PPN formalism}

Let us then begin by rewriting the metric tensor as a linear
combination of flat space and a metric perturbation $g_{ab} = \eta_{ab} + h_{ab}$.
Let us further work in the PPN formalism, where different components
of the metric perturbation are assumed to be of the following orders:
$h_{00} = {\cal{O}}(2)$, $h_{0i} = {\cal{O}}(3)$, $h_{ij} =
{\cal{O}}(2)$. In this section, the notation ${\cal{O}}(A)$ stands for
terms of order $\epsilon^A$, where $\epsilon$ is the
perturbation parameter of PN theory: the strength of the gravitational
field (i.~e.~an expansion in $G$) or the speed of particles (i.~e.~an
expansion in $1/c$). Note then that time derivatives are smaller by an
order of $\epsilon$ relative to spatial derivatives. We shall not review the PPN formalism in detail here, but instead we refer the reader to~\cite{Schiff:1960gi,Nordtvedt:1968qs,1972ApJ...177..775N,1971ApJ...163..611W,1973ApJ...185...31W,Will:1993ns}

We can now construct the Riemann tensor to leading order in the metric
perturbation. Let us restrict attention to a quasi-Cartesian
coordinate system, such that $\eta_{ab} dx^a dx^b = -dt^2 + dx^2 +
dy^2 + dz^2$. We then find that 
\be
R_{abc}{}^{d} = 2 \; \partial_{[a} \Gamma_{b] c}{}^{d} + {\cal{O}}(4),
\ee
and, following the conventions of~\cite{Romano:1991up}, 
$\Gamma_{ab}{}^{c} = -(1/2) \; g^{cd} \left[2 g_{d(a,b)} - g_{ab,d} \right]$ and then
\be
R_{abcd} = h_{d[a,b]}{}_c - h_{c[a,b]}{}_d + {\cal{O}}(4). 
\ee

With this linearized Riemann tensor we find that the CS torsion tensor becomes
\be
T_{cd}{}^n = \frac{1}{2}  \epsilon^{nbef} v_b 
h_{e[c,d]f} + {\cal{O}}(4).
\ee
Henceforth, we shall work to leading order in the torsion tensor and
consistently drop remainders of ${\cal{O}}(4)$ and higher. We can decompose
the torsion tensor into temporal and spatial components to find
\begin{subequations}
\ba
\label{st-term1}
T_{cd}{}^0 &=& \frac{1}{2} \epsilon^{0ijk} v_i h_{j[c,d]k},
\\
T_{cd}{}^i &=& - \frac{1}{2} \epsilon^{0ijk} v_0 h_{j[c,d]k}
+ \frac{1}{2} \epsilon^{0ijk} v_j \left(
  h_{0[c,d]k} - h_{k[c,d]0} \right),
\label{st-term2}
\nonumber \\
\ea
\end{subequations}
where we remind the reader that Latin indices in the middle of the alphabet $\{i,j,\ldots\}$ stand
for spatial indices only.  We recognize these terms as flat-space curls and cross
products of the metric perturbation and the CS velocity. Note that the CS acceleration does not
contribute to the torsion tensor.

Let us now specialize the torsion tensor to a specific source. In GR
and in the PPN formalism, the metric perturbation can be written to
first non-vanishing order as
\be
h_{00} = 2 U, \qquad
h_{0i} = - 4 V_i, \qquad
h_{ij} = 2 U \delta_{ij},
\ee
where $U$ is the Newtonian potential and $V_i$ is a PPN vector
potential. In general, the gravitomagnetic sector of the metric
contains two independent vector potentials, but in most cases of
interest, these vector potentials are identical. For example, for a
single stationary source at rest, these potentials are
\be
\label{ppn-potentials}
U = \frac{m}{r}, \qquad
V_i = \frac{m}{2 r^2} \tilde\eta_{ijk} a^j n^k,
\ee
where $m$ is the mass of the body, $a^i = J^i/m$ is the specific
angular momentum, $n^i = x^i/r$ is a unit vector and $r$ is the
distance from the center of the body to a field point. Note that since
$\dot{r} = 0$, all time derivatives of the metric vanish.

The metric perturbation presented above, however, is not a solution to
the CS modified field equations. For the case where $v_a =
(v_{0},0,0,0)$, such a solution can be constructed by adding a term
in the shift to the standard PPN
metric~\cite{Alexander:2007zg,Alexander:2007vt}:
\be
\delta h_{0i} = 2 \; v_{0} \; \tilde\eta_{ijk} \; V_{j,k}.
\ee
Note, however, that these CS corrections to $g_{ab}$ generically add
corrections to the torsion tensor that are at least quadratic in the
CS velocity, and thus, we shall neglect them.

We can now find all non-zero components of the torsion tensor in CS
modified gravity in terms of PPN potentials, namely:
\begin{subequations}
\ba
\label{simp-tor1}
T_{0l}{}^0 &=& - \epsilon^{0 i j k} \; v_i \; V_{j, k l},
\\
T_{ln}{}^0 &=&  \epsilon^{0 i k}{}_{[n} \; U_{,l ] k} \; v_i,
\\ \nonumber \\ 
T_{0l}{}^i &=& \epsilon^{0 i j k} \; v_0 \; V_{j, k l} + \frac{1}{2} 
\epsilon^{0 i j k} \; v_j \; U_{,l k},
\\ 
T_{ln}{}^i &=& \epsilon^{0 i j}{}_{ [l } \; U_{,n] j} \; v_0 - 2 v_j \epsilon^{0 i j k}
V_{ [l,n] k},
\label{simp-tor2}
\ea
\end{subequations}
where we have assumed the source is at rest and the CS velocity is
generic. The expressions presented above are generically valid for any
weakly-gravitating system in the far field.

\subsection{CS Torsion around Earth}

Let us now specialize the above analysis to bodies orbiting Earth and use 
the following line element
\ba
\label{earth-ds2}
ds^2 &=& \left(- 1 + \frac{2 M_{}}{{r}} \right) dt^2 + \left(1 +
  \frac{2 M_{}}{{r}} \right) dr^2 
\nonumber \\
&-& \frac{4 M_{} a}{r} \sin^2{\theta} \; dt \; d\phi, 
\ea
where $\{r,\theta,\phi\}$ are spherical polar coordinates. This line
element agrees both with the PPN formalism described above, as well as
with the far-field linearization of the Kerr line element. We shall
here treat both $M/r \ll 1$ and $a/r \ll 1$ as independent
perturbation parameters. For bodies orbiting Earth, these quantities
satisfy $M/r \sim {\cal{O}}(10^{-10})$ and $a/r \sim
{\cal{O}}(10^{-7})$.

We can now compute the components of the torsion tensor by inserting
the line-element of Eq.~\eqref{earth-ds2} into Eq.~\eqref{torsion}.
Alternatively, we could have used
Eqs.~\eqref{simp-tor1}-\eqref{simp-tor2} and the PPN potentials of the
previous section, but the calculation is simplified if instead of
quasi-Cartesian coordinates, we use spherical polar coordinates. We used 
{\emph{Maple}}~\cite{grtensor} to find that the largest non-vanishing components of the torsion
tensor are
\begin{subequations}
\ba
T_{r \theta}{}^t &=& -\frac{M}{2 r^3} \frac{v_{\phi}}{\sin{\theta}},
\\
T_{t \theta}{}^r &=& -\frac{M}{2 r^3 \sin{\theta}} \left(v_{\phi} + 3 v_t a
  \sin^2{\theta}  \right),
\\
T_{t \phi}{}^{\theta} &=& -\frac{M}{2 r^3} v_r \sin{\theta},
\\
T_{t \phi}{}^r &=& \frac{M}{2 r^3} v_{\theta} \sin{\theta},
\\
T_{t r}{}^{\theta} &=& - \frac{M}{2 r^5 \sin{\theta}} \left(2 v_{\phi} +
  3 v_t a  \sin^2{\theta} \right),
\\
T_{\theta \phi}{}^r &=& - \frac{M \sin{\theta}}{2 r^3} \left(2 v_t r^2 +
  3 v_{\phi} a \right),
\\
T_{r \phi}{}^{\theta} &=& - \frac{M \sin{\theta}}{2 r^5} \left( v_t r^2
  + 3 v_{\phi} a \right),
\\
T_{r \theta}{}^{\phi} &=& \frac{M}{2 r^3} \frac{v_t}{\sin{\theta}},
\ea
\end{subequations}
followed in magnitude by 
\begin{subequations}
\ba
T_{tr}{}^t &=& \frac{3 M a}{2 r^5} \left(2 v_r r \cos{\theta} +
  v_{\theta} \sin{\theta}\right),
\\
T_{t \theta}{}^t &=& \frac{3 M a}{2 r^4} \left(v_r r \sin{\theta} -
  v_{\theta} \cos{\theta} \right),
\\
T_{t \phi}{}^t &=& -\frac{3 M a}{2 r^4} v_{\phi} \cos{\theta},
\\
T_{t \phi}{}^{\phi} &=& \frac{3 M a}{2 r^4} v_t \cos{\theta},
\\
T_{r \theta}{}^r &=& -\frac{3 M a}{2 r^4} v_{\theta} \cos{\theta},
\\
T_{r \theta}{}^{\theta} &=& \frac{3 M a}{2 r^4} v_r \cos{\theta},
\\
T_{r \phi}{}^{\phi} &=& \frac{3 M a}{2 r^5} \left( v_r r \cos{\theta}
  + v_{\theta} \sin{\theta} \right),
\\
T_{\theta \phi}{}^{\phi} &=& \frac{3 M a}{2 r^4} \left(v_r
 r  \sin{\theta} - 2 v_{\theta} \cos{\theta} \right),
\ea
\end{subequations}
and all remaining components can be either obtained by symmetry or vanish to
this order. 

Torsion will affect geodesic motion through the Papapetrou equations,
which then leads to modified precession relative to the GR prediction.
We have checked that the only possible way for all components of the
torsion tensor to vanish is for the CS acceleration to identically
vanish $v_{a} = 0$. Even when $v_{r} = v_{\theta} = v_{\phi} = 0$, the
so-called canonical choice for $v_a$, there are still six
non-vanishing torsion tensor components. Clearly then, even for
non-canonical velocities where only $v_t = 0$, there are still ten
non-vanishing components of this tensor. We can conclude that
torsion and modified precession are inherent to CS modified gravity
irrespective of the choice of coupling parameter.

The torsion tensor presented here could be used to calculate the
change in the precessional angular frequency of gyroscopes orbiting
Earth.  Such a calculation is naturally interesting because it could
be compared to solar system experiments, such as Gravity Probe B, and
thus lead to a bound on the CS velocity. Such a test was first
proposed by Alexander and
Yunes~\cite{Alexander:2007zg,Alexander:2007vt}, who computed this
angular frequency in canonical CS modified gravity. Smith,
et.~al.~\cite{Smith:2007jm} extended this analysis and placed the first 
actual bounds
on the canonical choice of CS velocity. Unfortunately, such Solar
System bounds tend to be rather weak due to the feebleness of the
gravitational force in the Solar System.

The modified angular velocity of precession has not yet been
calculated for a generic choice of CS velocity. Mao,
et.~al.~\cite{Mao:2006bb} have considered Solar System tests of a
restricted class of torsion theories. This class is parameterized by
torsion tensors that are stationary, spherically or axially-symmetric
and parity preserving.  The torsion tensor associated with CS modified gravity
generically breaks parity unless one concocts a CS velocity that is
parity violating, such as the flat-space curl of some other field.
Even then, the explicit appearance of the CS velocity in the torsion
tensor tends to break spherical or axial symmetry. Thus, the torsion
tensor considered here is more general than that considered
in~\cite{Mao:2006bb}. Nonetheless, the analysis presented in this paper
provides a solid motivation for the study of the effect of generic torsion theories on 
Solar System experiments, similar in spirit to that of~\cite{Mao:2006bb}. A careful examination
of generic torsion theories and Solar System experiments is, however, 
beyond the scope of this paper.

\section{CS Modified Gravity and Fermions}
\label{Adding-Fermions} 

In this section, we study the inclusion of a minimally coupled fermion term to CS modified
lagrangian. Let us then consider the fermion-extended CS action
\begin{subequations}
\ba
S &=& S_{EH} + S_{CS} + S_{D}
\\
\label{Dirac-S}
S_{D} &=& \frac{1}{2} \int d^4x \sqrt{-g} \left(i \bar{\psi} \gamma^I
  e_I^a {\cal{D}}_a \psi + {\textrm{c.c.}} \right),
\ea
\end{subequations}
where the Einstein-Hilbert action $S_{EH}$ was given in
Eq.~\eqref{S_EH}, while the CS action $S_{CS}$ was presented in
Eq.~\eqref{CS-S}. Equation~\eqref{Dirac-S} is nothing but the massless Dirac
action~\footnote{The inclusion of a mass term does not affect the conclusions of this paper, but we leave it out to avoid cluttering.}, where ${\textrm{c.c.}}$ stands for complex conjugation, $\psi$
is a Dirac spinor, and $\gamma^I$ are gamma matrices. Note that the
first-order formalism is essential for the inclusion of fermions in
the theory, since Dirac spinors live naturally in $SU(2)$. Therefore,
covariant derivative associated with the Dirac action are not the
usual $SO(3,1)$ ones, but instead are given by
${\cal{D}}_a \psi := \partial_a \psi - (1/4) A_a{}^{IJ} \gamma_I
\gamma_J \; \psi$, where we here follow the sign conventions of~\cite{Perez:2005pm}.

Let us now find the field equations and the second structure equation of the fermion-extended CS modified gravity. The variation of the action with respect to the tetrad leads
to 
\ba
\tilde{\eta}^{abcd} \epsilon_{IJKL} e_b^J F_{cd}{}^{KL} = 8 \pi G T^a{}_I,
\ea
which is nothing but the Einstein equations with a generic connection
in the presence of a source,  given by Dirac fermions. 

The second structure equation can be obtained by varying the action
with respect to the connection. Following~\cite{Perez:2005pm} and using the identity 
\be
\gamma^{I} \gamma^{[J} \gamma^{K]} = - i \epsilon^{IJKL} \gamma_{5} \gamma_{L} + 2 \eta^{I[J} \gamma^{K]},
\ee
we find that the requirement that the variation of the action vanishes implies
\be
\label{torsion-eq1}
\tilde{\eta}^{abcd} \epsilon_{IJKL}  T_{cd}{}^e e_a^I e_b^J = 
    \tilde{\eta}^{ebcd} v_b  F_{cdKL} 
    - \frac{e}{\kappa} e^e_I \epsilon^I{}_{KLQ} \; J_5^Q 
\ee
where $J_5^Q := \bar{\psi} \gamma_5 \gamma^Q \psi$ is the axial
fermion current and where we the divergence of the tetrad vanishes. 
Once more, we can invert Eq.~\eqref{torsion-eq1} to leading order in the CS velocity to find an expression for the torsion: 
\be
\label{Torsion-w-fermions}
T_{cd}{}^n = - \frac{1}{4} \epsilon^{nbef} v_b R_{cdef} - \frac{1}{4
  \kappa} \epsilon^{n}{}_{cde} J^e_5 + {\cal{O}}(v)^{2}.
\ee

The torsion tensor can be manipulated slightly and written like a fermion term with a modified current. For this purpose, let us express the Riemann tensor in terms of its $4$-dimensional irreducible decomposition
\be
R_{abcd} = C_{abcd} + g_{a[c} R_{d]b} - g_{b[c} R_{d]a} - \frac{1}{3} R \; g_{a[c} g_{d]b},
\ee
where $C_{abcd}$ is the Weyl tensor. The torsion tensor can then be written as
\ba
T_{cd}{}^{n} &=& \frac{1}{12} \left(\epsilon^{nb}{}_{cd} v_{b} R + 6 \epsilon^{nbf}{}_{[c} R_{d]f} \right) - \frac{1}{4} \epsilon^{nbef} v_{b} C_{cdef} 
\nonumber \\
&-&  \frac{1}{4 \kappa} \epsilon^{n}{}_{cde} J_{5}^{e}.
\ea
For conformally flat spaces with constant curvature, we can use $R_{ab} = g_{ab} R/4$ and $C_{abcd} = 0$ to simplify the torsion into
\be
T_{cd}{}^{n} = - \frac{1}{4 \kappa} \epsilon^{n}{}_{cde} S_{5}^{e},
\ee
where we have defined the CS extended axial current
\be
S_{5}^{e} = J_{5}^{e} + \frac{\kappa}{6} v^{e} R.
\ee

The torsion tensor is related to the contorsion via Eq.~\ref{conteq1}, which can be inverted to find
\be
\label{def-cont}
C_{cd}{}^{n} = \frac{1}{2} \left[ T_{cd}{}^{n} + 2 T^{n}{}_{(cd)} \right].
\ee
One can check that the antisymmetrization of Eq.~\eqref{full-connection} in its first two indices with Eq.~\eqref{def-cont} for the contorsion returns the definition of torsion in Eq.~\eqref{def-torsion}~\footnote{The definition of the torsion and contorsion tensor are ambiguous up to a historical minus sign. The definition chosen in this paper have been checked to be internally consistent.}.
Using the torsion tensor found in Eq.~\eqref{Torsion-w-fermions}, the contorsion becomes
\ba
\label{comb-cont}
C_{cd}{}^{n} = - \frac{1}{8} \left[ {}^{\star}R_{cd}{}^{nb} v_{b} + 2 \; {}^{\star}R^{n}{}_{(cd)}{}^{b} v_{b} + \frac{1}{\kappa} \epsilon^{n}{}_{cde} J_{5}^{e} \right].
\ea
The fermionic part of the contorsion found here agrees with that found by~\cite{Perez:2005pm,Freidel:2005sn}
in the limit as the Immirzi parameter tends to infinity and the Holst action vanishes.

Once the torsion has been computed, one can reinsert it into the full action to obtain the equations of motion. We rewrite the full connection as the sum of a symmetric, torsion-free part $\omega_{a}{}^{bc}$ and the contorsion, as in Eq.~\eqref{full-connection}. Each contribution to the action then takes the form  $S = S[\omega] + S[C]$, where the first term is completely independent of the contorsion and the second term leads to contorsion-induced interaction terms. We find that contorsion-dependent contributions to the action are given by the following:
\be
S_{EH}[C] = \int d^{4}x \; e \left[ - \frac{1}{16} J_{5}^{b} \left(2 v^{a} R_{ab} - v_{b} R \right) 
+
 \frac{3}{32 \kappa} J_{5}^{a} J_{5 \; a} \right],
\ee
for the Einstein-Hilbert part;
\be
S_{D}[C] = \int d^{4}x \; e \left[ - \frac{1}{16} J_{5}^{b} \left[ -2 v^{a} R_{ab} + v_{b} R \right) 
-   \frac{3}{16 \kappa} J_{5}^{a} J_{5\;a} \right],
\ee
for the Dirac part; and
\ba
S_{CS}[C] &=& \int d^{4}x \; e \left[ \frac{v_{a}}{8} L^{a}[\omega \cdot \partial J,\; \omega \cdot \partial J \cdot \omega]
\right. 
\nonumber \\
&+& \left. 
\frac{1}{16} \left( J_{5}^{a} v_{a} R - 2 J_{5}^{a} v^{b} R_{ab} \right)
\right. 
\nonumber \\
&+& \left. 
 \frac{1}{64 \kappa} v_{a} \epsilon^{abcd} J_{5 d} \; \partial_{c} J_{5b}
\right. 
\nonumber \\
&-& \left.  \frac{1}{256 \kappa^{2}} \left(v_{a} J_{5}^{a}\right) \; \left(J_{5}^{b} J_{5 b} \right)
 \right],
\ea
for the CS part, where $L^{a}[\ldots]$ stands for a contractions of derivatives of the the axial current with the torsion-free connection. Interestingly, the CS contribution to the Einstein-Hilbert and Dirac parts of the action identically vanish, yielding the standard $J_{5}^{2}$ interaction of Riemann-Cartan theory in the presence of minimally coupled fermions:
\be
S_{EH}[C] + S_{D}[C] = - \frac{3}{2} \pi G \int d^{4}x \; e \; J_{5}^{a} J_{5 a}.
\ee
The CS contribution to the action, however, adds new parity violating interactions and the full action in Riemann normal coordinates ($\omega_{ac}{}^{d} = 0$) reduces to 
\ba
\label{full-S}
S[C] &=&  \int d^{4}x \; e \; \left[ \frac{1}{16} \left( J_{5}^{a} v_{a} R - 2 J_{5}^{a} v^{b} R_{ab} \right)
\right. 
\nonumber \\
&-& \left. 
 \frac{3}{2} \pi G \left( J_{5}^{a} J_{5 a} - \frac{1}{6}  \epsilon^{abcd} v_{a} J_{5 d} \; \partial_{c} J_{5b} \right) 
\right. 
\nonumber \\
&+&  \left. \pi^{2} G^{2} \left(v_{a} J_{5}^{a}\right) \; \left(J_{5}^{b} J_{5 b} \right) \right].
\ea

CS modified gravity has introduced a new parity-violating interaction that is not suppressed by Newton's gravitation constant and can be in fact enhanced by the CS velocity. This new interaction [first line in Eq.~\eqref{full-S}] is a two-fermion process that couples both to the CS velocity as well as to the spacetime curvature through the Ricci tensor and scalar. CS modified gravity also modifies the standard four-fermion interaction that is common to Riemann-Cartan theory with minimally coupled Dirac fermions. The modification consists of the addition of an antisymmetric product of the current and its derivative [second line in Eq.~\eqref{full-S}]. This interaction, however, is suppressed by one factor of $G$. Moreover, a new $6$-fermion interaction is produced, but this one is further suppressed by a factor of $G^2$.  

Before concluding, let us discuss one further equation of motion contained in fermion-extended CS modified gravity. Such an equation arises
because the Dirac action has an additional degree of freedom: the
Dirac spinor. Variation of the action with respect to the Dirac spinor
yields the Dirac equation (in this case, for a massless fermion), namely
$\gamma^a {\cal{D}}_a \psi = 0$. Splitting the connection into torsion-free and torsion-full pieces, 
one can easily show that the Dirac equation becomes
\be
\gamma^{a} D_{a}^{(\omega)} \psi = \frac{1}{4} e^{a}_{M} C_{a}{}^{KL} \gamma^{M} \gamma_{K} \gamma_{L} \psi
\label{Dirac-Eq},
\ee
where $D_{a}^{(\omega)}$ stands for the covariant derivative associated with the torsion-free connection only. 

The Dirac equation can then be thought of as {\emph{sourcing}} the contorsion tensor, 
and thus, the CS correction. One could insert the contorsion tensor found in 
Eq.~\eqref{comb-cont} to find an explicit equation for the relation between the torsion-free
Dirac equation and the fermion and CS torsion tensor. In essence, this equation implies that 
the CS effects will be enhanced in spacetime regions where the momentum of Dirac fermions 
is large. Several realistic astrophysical scenario exist where such an enhancement 
should be present, but we shall discuss these in the next section.

\section{Physical Implications}
\label{Discussion} 

In this section, we comment and theorize on some of the results found in the
previous sections and their consequences. As we have seen, CS modified
gravity can be mapped to a torsion theory, where the torsion tensor is
proportional to the CS velocity. Torsion then leads to a new and unsuppressed interaction term in the Dirac action, which represents a two-fermion process, mediated by the spacetime curvature and the CS velocity. 

Can such modified fermion interaction be detected in particle detectors? The four-fermion process is suppressed by a factor of the gravitational constant, so its detectability is questionable. The two-fermion interaction, however, is not suppressed by such a factor and it is precisely where the main CS correction resides. This interaction does depend on the Ricci tensor and scalar, which are close to zero in the neighborhood of the Solar System and might again suppress the effect. Nonetheless, such suppression might be overcome if the CS effect is enhanced by fermion currents. 

Are there any realistic physical scenarios where this enhancement
would actually occur? Such scenarios would require a large fermion current, which in essence implies large changes in fermion density. Dynamical compact stars, such as neutron stars and white dwarves, in fact possess a large fermion density, since they are supported against gravitational collapse via electron or neutron degeneracy pressure. Moreover, these systems can be dynamical, spinning rapidly, vibrating quasinormally, quaking or accreting mass from a binary companion.  Slightly more hypothetical sources, such as quark or strange stars~\cite{Freedman:1977gz,Alcock:1986hz}, would lead to even stronger enhancements since their fermion density is even larger.

Any fermionic compact object that undergoes
a violent change in its multipolar structure will also possess large
number density gradients. A few examples of such events are double neutron star or neutron star-black hole binary mergers and supernovae. In all such
systems, one of the binary components tidally disrupts and then either
collides or is swallowed by the black hole horizon, leading to a large
change in fermion number density. In the cosmological context, the big
bang event, as well as inflation, also unavoidingly lead to large fermion currents, where fermions accelerate violently.

Many of the sources described here are also target sources for
gravitational wave detection~\cite{Cutler:2002me}. Periodic
sources, such as pulsars, as well as inspiraling compact binaries of various types, are preferred systems for gravitational wave detection, because these are precisely produced by a changing multipolar structure. Moreover, the CS effect is not only enhanced by large fermion currents, but also for systems whose Riemann tensor becomes large. We then see that even
for binary black hole mergers, where there are no fermion currents in
play, the CS modification might play an important role. In this sense, the interplay between gravitational wave detection and CS modified gravity might be important in the future~\cite{Alexander:2007kv}. 

The CS modification to GR is naturally enhanced in a
plethora of realistic astrophysical scenarios. All that is required
for such an enhancement is the existence of large fermion currents, which are inherent in the
natural evolution of fermionic compact objects. We shall not quantify
the enhancement effect further in this paper, since this task is
extremely system dependent, and instead, we relegate it to future
work~\cite{pulsars}

\section{Conclusion}
\label{Conclusions} 

We have studied CS modified gravity in the tetrad
formalism. We rewrote the CS modified action in terms of a tetrad and
a generalized connection and varied it with respect to these fields to
obtain the modified field equations and the second structure equation. In doing so, we found that the
torsion-free condition of GR does not hold in CS modified gravity,
where now the torsion tensor is proportional to an antisymmetric product of
the Riemann tensor and derivatives of the scalar field.

We investigated this torsion tensor in the far field of a
weakly-gravitating object within the PPN framework. We found that the
torsion tensor is proportional to the contraction of
derivatives of the Newtonian and PPN vector potentials with the
Levi-Civita symbol and derivatives of the CS scalar. This torsion was
shown not to vanish for any non-trivial choice of the CS scalar, thus
suggesting that torsion is unavoidable in CS modified
gravity. Such torsion will generically lead to a modification of the
frame-dragging effect of GR, as well as a change in the precession of
gyroscopes. In this way, this paper provides great motivation for the
study of the effect of generic torsion theories on Solar System
experiments.

We then focused on the addition of fermions to the modified theory,
since these are known to also produce torsion when minimally coupled to Riemann-Cartan theory. We
indeed found that the torsion tensor was now composed of the CS
torsion piece, together with a new term that depends on the fermion
axial current. The torsion tensor was then used to reconstruct the full connection via the contorsion, which when inserted into the full action led to a new and unsuppressed two-fermion interaction. Although this interaction is not suppressed by the gravitation constant, it does depend on the Ricci tensor and scalar. This dependance might difficult its detection with particle detectors on Earth but might also enhance its effect in the neighborhood of compact sources. 

We then concentrated on the Dirac equation, which was shown to source
the CS torsion tensor. This enhancement effect depends both on the
fermion axial current and derivatives of the Dirac spinors, amplifying all CS effects when large changes in fermion density are present. 
We discussed the astrophysical implications of this enhancing
mechanism and found several interesting sources where such an
effect could be observed. Examples of such sources include binary neutron
star mergers, accreting neutron stars with their associated
supernovae, inflationary cosmology and the big bang. Not surprisingly,
we found that the enhancing mechanism seems to be maximized for the
same type of sources preferred by gravitational wave detection:
compact sources with large changes in their multipolar structure. 

Open questions still remain in the context of CS modified gravity. 
One such question is that of well-posedness as a  boundary
value problem. Although this issue has already been satisfactorily settled in~\cite{Grumiller:2008ie}
within the second order formalism, a similar study in the first order formalism is 
still absent. An analogous story must, of course, exist in first order form, and thus, 
one could construct the first order version of any required boundary terms and counter-terms
required by the variational principle to make the modified theory well-posed. 

Other future work could concentrate on studying the enhancing effect
further, in connection with some specific astrophysical scenario. For
example, one could study jet production in rapidly rotating neutron
stars or pulsars with the CS correction. Alternatively, one could
numerically investigate some simplified supernovae models in the
presence of a CS correction. 

Another possible avenue of future
research could be the numerical study of non-linear CS modified gravity.
For example, one could model the merger of binary black
holes in the modified theory to determine how waveforms change as a
function of the CS scalar. Perhaps most interesting is the numerical
evolution of the merger of black hole-neutron star systems, since
here both the fermionic enhancement and the curvature enhancements
should be present. Many of these open questions shall be answered in the near
future, hopefully shedding new light on some of the mysteries of the low-energy
limit of quantum gravity.

\acknowledgments We would like to thank Ben Owen for his continuous
support. We are also indebted to Daniel Grumiller and Andrew Randono, who 
provided essential help and clarifications with some calculations.
We would also like to thank  
Abhay Ashtekar, Roman Jackiw and Frans Pretorius
for useful comments and suggestions.   

SA acknowledges the support of the Eberly College of Science.
We both acknowledge the support of the Center for Gravitational Wave
Physics, which is funded by the National Science Foundation under
Cooperative Agreement PHY 01-14375. Some of the calculations presented in this paper were carried out with the symbolic manipulation software {\emph{Maple}}, in combination with the tensor package {\emph{GRTensorII}}~\cite{grtensor}.

\appendix
\section{Basics of the First-Order Formalism}

We here review the so-called {\emph{tetrad postulate}} following
Carroll~\cite{Carroll:2004st}. This postulate forces the covariant derivative
of the tetrad to vanish. Let us then begin by considering the
generalized covariant derivative of some vector $X$: 
\ba
{\cal{D}} X &=& \left({\cal{D}}_a X^b\right) dx^a \times \partial_b 
\nonumber \\
&=& \left(\partial_a X^b + A^b_{ab} X^c \right) dx^a \times \partial_b.
\ea
Let us now rewrite this quantity with internal indices, namely
\ba
{\cal{D}} X &=& \left({\cal{D}}_a X^I\right) dx^a \times \hat{e}_I,
\\ \nonumber 
&=& \left[ e_I^b X^d \partial_a e_d^I + e_I^b e_d^I \partial_a X^d +
  A_a{}^I{}_J e_d^J e^b_I X^d \right] dx^a \times \partial_b.
\ea
Comparing both expressions and requiring that they be equal we find
the constraint
\be
A_{ab}{}^c = e_I^c \partial_a e_b^I + e_b^J e_I^c A_a{}^I{}_J,
\ee
or equivalently
\be
\label{transf}
A_a{}^I{}_J = e_c^I A_{ab}{}^c \left(e_b^J\right)^{-1} -
\left(e_b^J\right)^{-1} \partial_a e_b^I. 
\ee
Equation\eqref{transf} shows clearly the character of the
transformation. From these equations, one can trivially derive that 
\be
{\cal{D}}_a e_b^I = 0.
\ee
Note that this relation was achieved without ever requiring metric
compatibility or torsion-freeness. Thus, the tetrad postulate is an
independent requirement that must always hold and unequivocally leads
to the vanishing of the covariant derivative of the tetrad.

The tetrad postulate, however, does not necessarily require metric
compatibility. One can easily show that 
\be
{\cal{D}}_a g_{bc} = e^I_b e^J_c {\cal{D}}_{a} \eta_{IJ},
\ee
which vanishes if and only if the connection is purely antisymmetric
on its internal indices, ie.~$A_{a(IJ)} = 0$. We see then that
spacetime metric compatibility is equivalent to internal metric
compatibility. We also clearly see that the tetrad postulate does not
automatically force metric compatibility. 

The generalized connection can be written in terms of the Christoffel
connection and the contorsion tensor. One can simply show that the metric
compatibility requirement $A_{a(IJ)} = 0$ together with the torsion
free condition $A_{(ab)c} = 0$ lead uniquely to $A_{ab c} =
\Gamma_{abc}$. When the torsion-free condition is dropped, this is not
the case anymore. In this case, the generalized connection is a linear
superposition of the Christoffel connection and the so-called
contorsion tensor, which can be constructed as linear superpositions
of the torsion tensor. 
\section{Torsion and Contorsion}
\label{app-cont}

In this appendix we review the definition of torsion and its relation to the contorsion
tensor, thus establishing further notation. Let us then consider first the torsion-free case, 
where the connection is simply given by $A_{a}{}^{IJ} = \omega_{a}{}^{IJ}$. The tetrad postulate then establishes that the transformation between the spin and the spacetime connection 
\be
\omega_{a}{}^{IK} e^{K}_{b} = \partial_{a} e_{b}^{I} + \omega_{ab}{}^{\delta} e_{\delta}^{I}.
\ee
Furthermore, the antisymmetrization of the tetrad postulate on the spacetime indices yields
\be
\label{torsion-free-transf}
\partial_{[a} e_{b]}^{I} = \omega_{a}{}^{IK} e_{b}^{K},
\ee
where $\omega_{[ab]}{}^{\delta} = 0$ by the torsion-free condition. Equation~\eqref{torsion-free-transf} establishes that the torsion-free spin connection is nothing but the transformed Christoffel connection. 

Let us now consider the torsion-full case, where we split the full connection $A_{a}{}^{IJ}$ into a torsion-free piece $\omega_{a}{}^{IJ}$ and a torsion-full part $C_{a}{}^{IJ}$. We further assume that the torsion-free piece satisfies the tetrad postulate with respect to the torsion-free covariant derivative, thus once more rendering $\omega_{a}{}^{IJ}$ the transformed Christoffel connection. The tetrad postulate with respect to the full covariant derivative yields the transformation law for the contorsion
\be
\label{torsion-full-transf}
C_{ab}{}^{c} e_{c}^{I} = C_{a}{}^{IK} e_{bK}.
\ee
Antisymmetrizing the lower two spacetime indices in Eq.~\eqref{torsion-full-transf} and using the definition of torsion $T_{ab}{}^{c} := 2 A_{[ab]}{}^{c}$ leads to the relation between contorsion and torsion:
\be
T_{ab}{}^{c} e_{c}^{I} = 2 C_{[a}{}^{IK} e_{b]K}.
\ee
The inversion of the torsion-contorsion relation then yields explicitly Eq.~\eqref{def-cont}. 



\end{document}